\documentclass[amsmath,amssymb,11pt,superscriptaddress,reprint, preprintnumbers, notitlepage,aps,prl,twocolumn, dvipsnames,table]{revtex4-1}
\pdfoutput=1 
\usepackage[utf8]{inputenc}
\usepackage[english]{babel}
\usepackage{amsmath}
\usepackage{graphicx}
\usepackage{dcolumn}
\usepackage{pbox}
\usepackage{amssymb}
\usepackage{epsfig}
\usepackage[normalem]{ulem}
\usepackage{slashed}
\usepackage{amssymb}
\usepackage{verbatim} %for comment environment
\usepackage{ mathrsfs }
\usepackage{color}
\usepackage[font=small]{caption}
\usepackage[font=small]{subcaption}
\usepackage{url}
\definecolor{MyDarkBlue}{rgb}{0.1, 0.1, 0.8}
\definecolor{SBlue}{rgb}{0.2, 0.4, 0.7} 
\definecolor{MyLightBlue}{rgb}{0.22,0.51,0.9}
\definecolor{MyGreen}{rgb}{0.0, 0.5, 0.0}
\definecolor{BrickRed}{rgb}{0.8, 0.25, 0.33}
\RequirePackage{hyperref}
\hypersetup{colorlinks, citecolor=SBlue,linkcolor=MyDarkBlue, urlcolor=PineGreen}
\usepackage[dvipsnames]{xcolor}

\usepackage{tikz}
\usepackage{multirow}
\usepackage{colortbl}

\usepackage{amsmath}
\usepackage{amssymb}
\usepackage{array}
\usepackage{multirow}
\usepackage{booktabs}

\makeatletter

\makeatletter
\renewcommand\@makecaption[2]{%
  \par
  \vskip\abovecaptionskip
  \begingroup
  
   \small\rmfamily
    \begingroup
     \samepage
     \flushing
     \let\footnote\@footnotemark@gobble
     \@make@capt@title{#1}{#2}\par
    \endgroup
  \endgroup
  \vskip\belowcaptionskip
}
\makeatother

\DeclareUnicodeCharacter{2212}{-}
\begin{document}

\preprint{MS-TP-25-13}
%%%%%%%%%%%%%%%%%%%%%%%%%%%%%%
\title{\vspace{1cm}\large 
How Charged Can Neutrinos Be? 
}

%%%%%%%%%%%%%%%%%%%%%%%%%%%%%%       
\author{\bf Sudip Jana}
\email[E-mail:]{sudip.jana@okstate.edu}
\affiliation{Harish-Chandra Research Institute, A CI of Homi Bhabha National Institute, Chhatnag Road, Jhunsi, Prayagraj 211 019, India}
%%%%%%%%%%%%%%%%%%%%%%%%%%%%%%
\author{\bf Michael Klasen}
\email[E-mail:]{michael.klasen@uni-muenster.de}
\affiliation{Institut f{\"u}r Theoretische Physik, Universit{\"a}t M{\"u}nster, Wilhelm-Klemm-Stra\ss{}e 9, 48149 M{\"u}nster, Germany}
%%%%%%%%%%%%%%%%%%%%%%%%%%%%%%
\author{\bf Vishnu P.K.}
\email[E-mail:]{vishnu.pk@uni-muenster.de}
\affiliation{Institut f{\"u}r Theoretische Physik, Universit{\"a}t M{\"u}nster, Wilhelm-Klemm-Stra\ss{}e 9, 48149 M{\"u}nster, Germany}
%%%%%%%%%%%%%%%%%%%%%%%%%%%%%%
%%%%%%%%%%%%%%%%%%%%%%%%%%%%%%

\begin{abstract}
We investigate how neutrinos may acquire small electric charges within the Standard Model framework while preserving electromagnetic gauge invariance. Instead of gauging the standard hypercharge generator \( Y \), a linear combination of \( Y \) and a new generator \( X \) from a gaugable global \( U(1)_X \) symmetry is embedded, under which neutrinos transform non-trivially. We demonstrate that minimal scenarios based on flavor-dependent \( U(1)_X \) symmetries, such as \( X = L_\alpha - L_\beta \), are incompatible with current neutrino oscillation data. In contrast, we have shown that only flavor-universal \( U(1)_X \) symmetries—such as \( U(1)_{B-L} \), which shifts both quark and lepton charges, and \( U(1)_L \), which modifies only the lepton sector—can generate tiny neutrino charges consistent with observed masses and mixing. We also discuss the necessary connection between such charges and the Dirac nature of neutrinos. By analyzing the phenomenological implications in detail, our findings emphasize that constraints on neutrino charges should be evaluated within the specific framework of the \( U(1)_X \) symmetry under consideration, rather than assuming a generic approach, as is often the case.
\end{abstract}

\maketitle
%%%%%%%%%%%%%%%%%%%%%%%%%%%%%%%%%%%%%%%%%%%%%%%%%%
%%%%%%%%main text%%%%%%%%%%%%%%%%%%%%%%%%%%%%%%%%%

\textbf{\emph{Introduction}.--}
The quantization of electric charge---where all known elementary particles possess charges that are integer multiples of one-third the electron charge---remains one of the unresolved mysteries in particle physics. While gauge invariance enforces the neutrality of gauge bosons ($Z$ boson, gluons, photon), no equally fundamental principle mandates that neutrinos be strictly electrically neutral. The possibility that neutrinos might deviate from exact charge neutrality has been a recurring theme in theoretical physics over the past decades, especially in the context of understanding charge quantization \cite{Babu:1989tq, Foot:1998tb}. Beyond its theoretical significance, even a tiny electric charge for neutrinos can have far-reaching astrophysical, cosmological, and experimental implications. For instance, during the early attempts to resolve the solar neutrino problem, multiple solutions were proposed---including neutrino oscillations enhanced via the Mikheyev-Smirnov-Wolfenstein (MSW) mechanism, neutrino decay, spin precession in solar magnetic fields, and resonant spin-flavor conversions. An intriguing and unconventional possibility suggests that if neutrinos possess even a minuscule electric charge, their trajectories could be altered by the Sun’s magnetic field due to the Lorentz force, leading to an anisotropic solar neutrino flux \cite{Ignatiev:1994fd}. Such deviations would cause the solar neutrino flux to become directionally dependent, potentially resulting in a reduced neutrino count detected on Earth compared to the expectations of the Standard Solar Model. This mechanism offered a compelling explanation independent of oscillations, spin flips, or decay. Although the MSW effect \cite{Wolfenstein:1977ue, Mikheev:1986wj, Mikheyev:1985zog} is now well established as the primary solution to the solar neutrino problem \cite{Maltoni:2015kca}, the idea of charged neutrinos has not lost relevance. Even an extremely small electric charge could significantly influence solar, atmospheric, accelerator-based, relic, and high-energy astrophysical neutrino observations. In the coming decades, numerous direct dark matter detection and neutrino scattering experiments---both ongoing and planned---will offer unprecedented sensitivity. If neutrinos carry a tiny electric charge, this could lead to observable implications in laboratory like nuclear or electron recoil excesses in such experiments.

In the Standard Model (SM), neutrinos are electrically neutral and do not couple to photons at tree level, interacting solely through weak interactions mediated by the \( W \) and \( Z \) bosons. Nevertheless, in various extensions of the SM, neutrinos can acquire effective electromagnetic interactions---such as magnetic moments, charge radii, or electric dipole moments---via quantum loop corrections \cite{Giunti:2014ixa, Babu:2020ivd, Babu:2021jnu, Bernabeu:2000hf}. If neutrinos were to possess even a minute electric charge, they would directly couple to photons at tree level, leading to significant phenomenological consequences. Interestingly, the SM does not inherently forbid neutrinos from carrying small electric charges, provided that anomaly cancellation conditions remain satisfied. This naturally motivates the question: \textit{What happens if neutrinos aren't completely neutral?} Addressing this requires a careful analysis of the SM's global symmetry structure. In addition to the gauged hypercharge symmetry \( U(1)_Y \), the SM includes four global \( U(1) \) symmetries---\( U(1)_B \) and \( U(1)_{L_e}, U(1)_{L_\mu}, U(1)_{L_\tau} \)---associated with baryon and individual lepton flavor numbers. These symmetries are accidental, emerging from the structure of the Lagrangian, and are not associated with any gauge fields.

It is crucial to observe that \( U(1)_Y \) is not the only anomaly-free Abelian symmetry within the SM. Instead of gauging only the standard hypercharge, one could consider gauging a linear combination of \( U(1)_Y \) and an additional anomaly-free global symmetry \( U(1)_X \), under which neutrinos transform non-trivially. This approach naturally gives rise to charged neutrinos. Most of the earlier literature discussing neutrino electric charges has centered around family-dependent lepton number differences, such as \( U(1)_{L_\mu - L_\tau} \)~\cite{Foot:1990mn, Babu:1992sw}, which are also anomaly-free within the SM. However, we demonstrate that minimal models, where neutrinos acquire electric charges through flavor-dependent \( U(1)_X \) symmetries are incompatible with the observed neutrino oscillation data, including scenarios based on \( U(1)_{L_\mu - L_\tau} \). This is because such models fail to correctly generate the neutrino masses and mixing angles as predicted by current oscillation experiments. In contrast, flavor-universal \( U(1)_X \) symmetries offer viable solutions. We present and propose two specific examples: \( U(1)_X = U(1)_{B-L} \) and \( U(1)_X = U(1)_L \). In the first, both quark and lepton electric charges are modified, while in the second, only the lepton charges are altered. We explore the detailed phenomenological consequences of these frameworks, showing that they not only preserve anomaly cancellation but also remain consistent with current experimental constraints. We also highlight that the constraints on neutrino charges cannot be treated generically---as is often assumed in the literature~\cite{Giunti:2014ixa, Giunti:2024gec}---but must instead be evaluated within the specific context of the underlying \( U(1)_X \) symmetry.

In the following section, we explore the potential connections between the Dirac nature of neutrinos and the scenario of charged neutrinos. We then analyze pathways to mini-charged neutrinos in both flavor-dependent \( U(1)_X \) and flavor-universal \( U(1)_X \) scenarios. We also investigate the possibility of embedding charged neutrinos within non-Abelian gauge extensions, focusing on an example from the left-right symmetric theory. Subsequently, we examine the phenomenological implications of each case in detail. Our findings highlight that constraints on neutrino charges should be assessed within the specific context of the \( U(1)_X \) symmetry considered, rather than adopting a generic approach, as is commonly done. Finally, we conclude.

\begin{table*}[t!]
\centering
\renewcommand{\arraystretch}{1.2}
\setlength{\tabcolsep}{0pt}
\scriptsize
\arrayrulecolor{black}
\begin{tabular}{|c|c|c|c|}
\hline
\rowcolor{gray!30} \textbf{Serial} & \textbf{Gauge} & \textbf{Electric charges of} & \textbf{Consistent with} \\
\rowcolor{gray!30} \textbf{No.} & \textbf{Symmetry} & \textbf{the Matter Fields} & \textbf{$\nu$ oscillation data?}\\
\hline
\rowcolor{red!10}  &  & $Q_u = \frac{2}{3},\ Q_d = -\frac{1}{3}$ &  \\
\rowcolor{red!10}  1  & $SU(3)_C \times SU(2)_L \times U(1)_Y$ & $Q_e = -1,\ Q_\mu = -1,\ Q_\tau = -1$ & \\
\rowcolor{red!10}  &  & $Q_{\nu_e} = 0,\ Q_{\nu_\mu} = 0,\ Q_{\nu_\tau} = 0$ & \\
\hline
\rowcolor{blue!10}  &  & $Q_u = \frac{2}{3},\ Q_d = -\frac{1}{3}$ & \\
\rowcolor{blue!10} 2 & $SU(3)_C \times SU(2)_L \times U(1)_{Y + \epsilon(L_e - L_\mu)}$ & $Q_e = -1 + \epsilon_\nu,\ Q_\mu = -1 - \epsilon_\nu,\ Q_\tau = -1$ & \huge\textcolor{purple}{$\times$}\\
\rowcolor{blue!10}  &  & $Q_{\nu_e} = +\epsilon_\nu,\ Q_{\nu_\mu} = -\epsilon_\nu,\ Q_{\nu_\tau} = 0$ & \\
\hline
\rowcolor{yellow!10}  &  & $Q_u = \frac{2}{3},\ Q_d = -\frac{1}{3}$ &  \\
\rowcolor{yellow!10} 3 & $SU(3)_C \times SU(2)_L \times U(1)_{Y + \epsilon(L_e - L_\tau)}$ & $Q_e = -1 + \epsilon_\nu,\ Q_\mu = -1 ,\ Q_\tau = -1 + \epsilon_\nu$ & \huge\textcolor{purple}{$\times$}\\
\rowcolor{yellow!10}  &  & $Q_{\nu_e} = +\epsilon_\nu,\ Q_{\nu_\mu} = 0,\ Q_{\nu_\tau} = -\epsilon_\nu$ & \\
\hline
\rowcolor{green!10}  &  & $Q_u = \frac{2}{3},\ Q_d = -\frac{1}{3}$ & \\
\rowcolor{green!10} 4 & $SU(3)_C \times SU(2)_L \times U(1)_{Y + \epsilon(L_\mu - L_\tau)}$ & $Q_e = -1,\ Q_\mu = -1 + \epsilon_\nu,\ Q_\tau = -1 - \epsilon_\nu$ & \huge\textcolor{purple}{$\times$}\\
\rowcolor{green!10}  &  & $Q_{\nu_e} = 0,\ Q_{\nu_\mu} = \epsilon_\nu,\ Q_{\nu_\tau} = -\epsilon_\nu$ & \\
\hline
\rowcolor{orange!10}  &  & $Q_u = \frac{2}{3} - \frac{\epsilon_\nu}{3},\ Q_d = -\frac{1}{3} - \frac{\epsilon_\nu}{3}$ &  \\
\rowcolor{orange!10} 5 & $SU(3)_C \times SU(2)_L \times U(1)_{Y + \epsilon(B - L)}$ & $Q_e = -1 + \epsilon_\nu,\ Q_\mu = -1 + \epsilon_\nu,\ Q_\tau = -1 + \epsilon_\nu$ & \huge\textcolor{teal}{$\checkmark$}\\
\rowcolor{orange!10}  &  & $Q_{\nu_e} = +\epsilon_\nu,\ Q_{\nu_\mu} = +\epsilon_\nu,\ Q_{\nu_\tau} = +\epsilon_\nu$ & \\
\hline
\rowcolor{purple!10}  &  & $Q_u = \frac{2}{3},\ Q_d = -\frac{1}{3}$ &  \\
\rowcolor{purple!10} 6 & $SU(3)_C \times SU(2)_L \times U(1)_{Y + \epsilon(L)}$ & $Q_e = -1 + \epsilon_\nu,\ Q_\mu = -1 + \epsilon_\nu,\ Q_\tau = -1 + \epsilon_\nu$ & \huge\textcolor{teal}{$\checkmark$}\\
\rowcolor{purple!10}  &  & $Q_{\nu_e} = +\epsilon_\nu,\ Q_{\nu_\mu} = +\epsilon_\nu,\ Q_{\nu_\tau} = +\epsilon_\nu$ & \\
\hline
\end{tabular}
\caption{Modification of gauge symmetries and the resulting charges of matter fields to accommodate charged neutrinos.}
\label{tab:gauge_symmetry}
\end{table*}

\textbf{\emph{Dirac nature of charged neutrinos}.--}
In general, neutrinos can be either Dirac or Majorana in nature. However, in the case of charged neutrinos, the Majorana mass term, which breaks the lepton number by two units, is forbidden by the electromagnetic gauge symmetry, and hence they are Dirac type in nature. As a consequence of this, building any models of charged neutrinos would inherently be a realization of Dirac neutrinos.  

This could be compared with other realizations of Dirac neutrinos, see refs.~\cite{Lee:1956qn, Foot:1991py, Berezhiani:1995yi, Silagadze:1995tr, Farzan:2012sa, Ma:2014qra, Ma:2015mjd, CentellesChulia:2016rms, Bonilla:2016diq, Ma:2016mwh, Wang:2016lve, Borah:2017dmk, Jana:2019mez, Gu:2019ogb, Jana:2019mgj}. In these scenarios,  one requires at least an additional symmetry to protect the Dirac nature of neutrinos. Moreover, if this symmetry is off, discrete or global in nature, then the Planck suppressed higher dimensional operators could lead to Majorana mass terms for  neutrinos, yielding pseudo-Dirac neutrinos in effect \cite{Wolfenstein:1981kw, Petcov:1982ya, Valle:1983dk, Kobayashi:2000md, Babu:2022ikf, Biswas:2024wbz}. On the other hand, in the case of charged neutrinos, their Dirac nature is protected by the electromagnetic gauge symmetry and hence no additional beyond the SM (BSM) symmetry is required. Furthermore, since the Planck suppressed operators are expected to respect gauge symmetry, their Dirac nature is protected to all orders.

%%%%%%%%%%%%%%%%%%%%%%%%%%%%%%%%%%%%%%%%%%%%%%%%%%%%%%
%%%%%%%%%%%%%%%%%%%%%%%%%%%%%%%%%%%%%%%%%%%%%%%%%%%%%%
%\medskip
\textbf{\emph{Pathways to mini-charged neutrinos}--} 
%\section{Pathways to minicharged neutrinos}
%%%%%%%%%%%%%%%%%%%%%%%%%%%%%%%%%%%%%%%%%%%%%%%%%%%%%%
%%%%%%%%%%%%%%%%%%%%%%%%%%%%%%%%%%%%%%%%%%%%%%%%%%%%%%
Here we outline the method we followed to build models of charged neutrinos. 
Our approach is based on the electric charge dequantization property of neutrino mini-charged theories. In general, electric charge dequantization arises when (1) the electromagnetic gauge symmetry is broken or (2) the theory holds at least one gaugable global symmetry, which is not the same as the SM hypercharge symmetry \cite{Foot:1989fh, Foot:1990mn, Babu:1992sw, Foot:1992ui}. In this work, we exploit the second scenario to construct theories of charged neutrinos \footnote{See Refs.~\cite{Maruno:1991vr, Takasugi:1991ai, Takasugi:1991wa, Ignatiev:1996np, Ignatiev:1997pk, Babu:1990vw} for scenarios of charged neutrinos emerging from the breaking of electromagnetic gauge symmetry.}

A crucial ingredient of this scheme is models that possess a gaugable global symmetry $U(1)_X$ under which neutrino transforms non-trivially (the corresponding quantum number is denoted as $X_{\nu}$). For concreteness, we focus on the models based on the SM gauge symmetry. In this scenario, instead of gauging the SM hypercharge generator $Y$, one can gauge a linear combination of $Y$ and $X$, where $X$ represents the generator of $U(1)_X$ symmetry. The corresponding altered hypercharge symmetry is denoted as $U(1)_{Y+\epsilon X}$, where $\epsilon$ is a free parameter. Then, the spontaneous symmetry breaking of the modified electroweak gauge group $SU(2)_L\times U(1)_{Y+\epsilon X}$ yields an unbroken electromagnetic symmetry  $U(1)_{Q}$ with neutrinos of charge $\epsilon X_{\nu}$.  

To implement this method successfully, the global symmetry $U(1)_X$ has to satisfy the following requirements:
\begin{enumerate}
    \item $U(1)_X$ symmetry has to comply with all the gauge anomaly conditions.
    \item $U(1)_X$ symmetry is neither explicitly nor spontaneously broken. 
    \item Under $U(1)_X$ symmetry, the SM leptons should transforms non-trivially.
\end{enumerate}
Conditions 1 and 2 ensure that 
any linear combination of $Y$ and $X$ can be a gaugable symmetry of the model and 
also guarantee that after the symmetry breaking of $SU(2)_L\times U(1)_{Y+\epsilon X}$, the electric charge of the particles will be modified in the following way
\begin{align}
    Q=Q_{st} + \epsilon X.
\end{align}
Here $Q_{st}$  is the standard electric charge of the particle, given by $Q_{st}=I_3 + Y$, where $I_3$ represents the third component of $SU(2)_L$ symmetry. Condition 3 assures that neutrinos acquire a non-zero electric charge.

%%%%%%%%%%%%%%%%%%%%%%%%%%%%%%%%%%%%%%%%%%%%%%%%%%%%%%
%%%%%%%%%%%%%%%%%%%%%%%%%%%%%%%%%%%%%%%%%%%%%%%%%%%%%%
\textbf{\emph{Models of mini-charged neutrinos}--} 
%%%%%%%%%%%%%%%%%%%%%%%%%%%%%%%%%%%%%%%%%%%%%%%%%%%%%%
%%%%%%%%%%%%%%%%%%%%%%%%%%%%%%%%%%%%%%%%%%%%%%%%%%%%%%
With this strategy, we systematically look for models of charged neutrinos originating from various scenarios of gaugable $U(1)_X$ symmetry. For this, we classify the models that possess anomaly-free $U(1)_X$ symmetry into two categories: (1) flavor-dependent $U(1)_X$ scenarios and (2) flavor-universal $U(1)_X$ scenarios. 
%%%%%%%%%%%%%%%%%%%%%%%%%%%%%%%%%%%%%%%%%%%%%
\subsection{Flavor dependent $U(1)_X$}
%%%%%%%%%%%%%%%%%%%%%%%%%%%%%%%%%%%%%%%%%%%%$$
First, we consider the flavor-dependent $U(1)_X$ scenarios, wherein the different flavors of the SM have different charges under the $U(1)_X$ symmetry. There are several models in this category. Among them, we consider $U(1)_{L_i-L_j}$ and $U(1)_{B_i-L_i}$ as prototypical models for our study, where $i$ and $j$ are flavor indices ($i,j=1-3$). We also comment on various other models of this category at the end.

The $U(1)_{L_i-L_j}$ symmetry is anomaly free without any need to extending the SM particle sector~\cite{Foot:1989fh,He:1990pn}. On the other hand, the $U(1)_{B_i-L_i}$ case would requires a right-handed neutrino field $\nu_R$ (with quantum number $-1$ under $U(1)_{B_i-L_i}$ symmetry) to satisfy the gauge anomaly conditions. Moreover, in order to ensure condition 2, the Majorana mass term for $\nu_R$ is not allowed. 
With these considerations, both scenarios satisfy all the three necessary conditions that we have mentioned in the last section. For concreteness, we choose $i=3$ and $j=2$.
By following the aforementioned strategy, we gauge a linear combination of $Y$ and $L_{\tau}-L_{\mu}$ ($B_{3}-L_{3}$ for $U(1)_{B_{3}-L_{3}}$ scenario), which leads to 
\begin{align}
    &Q_{\nu_{\mu}}=-\epsilon,  \quad Q_{\nu_{\tau}}=\epsilon \notag \\
    &Q_{\mu}=-1-\epsilon,  \quad Q_{\tau}=-1 +\epsilon
\end{align}
for $U(1)_{L_{\tau}-L_{\mu}}$ scenario~\cite{Foot:1990mn, Babu:1992sw} and 
\begin{align}
    &Q_{\nu_{\tau}}=-\epsilon,  \quad Q_{\tau}=-1-\epsilon \notag \\
    &Q_{t}=\frac{2}{3}+\frac{\epsilon}{3},  \quad Q_{b}=-\frac{1}{3}+\frac{\epsilon}{3}
\end{align}
for $U(1)_{B_{3}-L_{3}}$ scenario~\cite{Foot:1990mn}. The charges of the other SM particles are unaltered with respect to the standard scenario. 

Despite the success in accommodating the charged neutrinos, these scenarios are not compatible with various experimental data. For instance, $U(1)_{L_i-L_j}$ scenario is not consistent with the neutrino oscillation data, since the mixing between the different flavors of neutrinos are forbidden due to the difference in their electric charges (as a consequence of unbroken  $U(1)_{L_i-L_j}$ symmetry). Additionally, two neutrinos are also massless in this framework. On the other hand, in the case of $U(1)_{B_i-L_i}$, in addition to these incompatibilities, it is also inconsistent with the observed quark mixings. In general, other flavor dependent $U(1)_X$ scenarios also suffer from such inconsistencies with experimental data. 
 
%%%%%%%%%%%%%%%%%%%%%%%%%%%%%%%%%%%%%%%%%%%%%
\subsection{Flavor universal $U(1)_X$}
%%%%%%%%%%%%%%%%%%%%%%%%%%%%%%%%%%%%%%%%%%%%$$
Next, we consider the flavor universal $U(1)_X$ scenarios, in which the SM flavors have the same charge under the $U(1)_X$ symmetry. 
This category includes cases such as $U(1)_{B-L}$ and $U(1)_L$. 

First, we consider the $U(1)_{B-L}$ scenario. In this case, the SM needs to be extended with three right-handed neutrinos (with $-1$ charge under $U(1)_{B-L}$ symmetry) to satisfy the gauge anomaly conditions. Similar to the $U(1)_{B_i-L_i}$ case, in order to comply with the condition 2, the Majorana mass terms for right-handed neutrinos are not allowed. Moreover, neutrinos are charged under this symmetry. Hence, this scenario satisfy all the three necessary conditions required to build a model of charged neutrinos. By following our strategy, we gauge a linear combination of $Y$ and $B-L$, which yields \cite{Foot:1992ui}
\begin{align}
    Q=Q_{st} + \epsilon (B-L),
\end{align}
after the spontaneous symmetry breaking of modified EW-gauge group. This implies that, in this setup, neutrinos will have a non-zero electric charge $-\epsilon$. Moreover, the electric charges of other SM particles are also altered, which can be explicitly written as 
\begin{align}
    Q_u= \frac{2}{3} +  \frac{\epsilon}{3}, \quad Q_d= -\frac{1}{3} + \frac{\epsilon}{3}, \quad Q_e= -1 - \epsilon. 
\end{align}
Here $u,d$, and $e$ represent the up-type quarks, down-type quarks, and charged leptons, respectively. Note that the electric charge of  particles is modified in a flavor universal way. This implies quark mixings and neutrino mixings are allowed in this framework. The neutrino masses are generated via the Yukawa interactions
\begin{align}
    \mathcal{L}_Y \supset Y_{\nu} \overline{\ell_L} \widetilde{H} \nu_R +h.c.
\end{align}
Here the fields $\ell_L$ and $H$ ($\widetilde{H}\equiv i\sigma_2H^*$) denote the left-handed lepton doublets and the Higgs-doublet, respectively.

Next, we turn our attention to the $U(1)_L$ scenario. Similar to the $U(1)_{B-L}$ setup, in this case the SM particle sector must be extended in order to comply with the gauge anomaly conditions. Here, we consider one such possibility~\cite{Chao:2010mp}, which contains the following BSM states 
\begin{align}
&\nu_{R_i}\sim(1,0,1), \quad i=1-3, \nonumber \\
&\psi^{1,2}_L = \begin{pmatrix} \psi_1^{1,2}  \\ \psi_2^{1,2}  \end{pmatrix}_L\sim (2,\pm a, -\dfrac{3}{2}),  \nonumber\\ 
%& \psi^2_L = \begin{pmatrix} \psi_1^{2}  \\ \psi_2^{2}  \end{pmatrix}_L\sim (2,-a, -\dfrac{3}{2})  \nonumber\\ 
&\psi^{1,2}_{1R}\sim(1,\pm a+\dfrac{1}{2},-\dfrac{3}{2}), \quad \psi^{1,2}_{2R}\sim(1,\pm a-\dfrac{1}{2},-\dfrac{3}{2}),
\end{align}
where each parenthesis contains quantum numbers of the corresponding particle under $\{SU(2)_L, U(1)_Y, U(1)_L\}$ symmetries. Here, $a$ can take any value; for concreteness, we choose $a=1/2$. With this particle content, the most general Yukawa interactions which respect  $\{SU(3)_C, SU(2)_L, U(1)_Y, U(1)_L\}$ symmetries can be written as
\begin{align}
\mathcal{L}_{\rm Yuk} =&  \mathcal{L}_Y^{SM} + Y_{\nu} \overline{\ell_L} \widetilde{H} \nu_R + Y_1 \overline{\psi^1_{L} }\psi_{1R}^1  \widetilde{H} + Y_2 \overline{\psi^2_{L} } \psi_{2R}^2  H \nonumber\\ 
& + Y_3 \overline{\psi^1_{L} } \psi_{2R}^1  H + Y_4 \overline{\psi^2_{L} }\psi_{1R}^2  \widetilde{H} + Y_5 \overline{\psi^1_{L} } \psi_{1R}^2  H \nonumber\\ 
&+ Y_6 \overline{\psi^2_{L} }\psi_{2R}^1  \widetilde{H} + h.c. 
\end{align}
Here $\mathcal{L}_Y^{SM}$ is the SM Yukawa Lagrangian. 
As noted earlier, the Majorana mass terms  $M_{\nu}\nu_R \nu_R$ are forbidden by $U(1)_L$ symmetry. This implies that the neutrino masses solely generated via the Yukawa couplings $Y_{\nu}$. On the other hand, the masses of all BSM states are generated via the Yukawa couplings $Y_{1-6}$.

The gauging of linear combination of $Y$ and $L$ leads to altering the electric charges of the particles in the following way 
\begin{align}
    Q=Q_{st} + \epsilon L.
\end{align}
It is evident from the above equation that  neutrinos have a non-zero electric charge $\epsilon$. Moreover,  the electric charges of the SM charged leptons are also altered, $Q_e=-1+\epsilon$. 
However,  charges of the SM quark fields are not modified with respect to the standard scenario in this framework.

%%%%%%%%%%%%%%%%%%%%%%%%%%%%%%%%%%%%%%%%%%%%%%%%%%%%%%
%%%%%%%%%%%%%%%%%%%%%%%%%%%%%%%%%%%%%%%%%%%%%%%%%%%%%%
\textbf{\emph{Mini-charged neutrinos within a left-right symmetric model}--} 
%%%%%%%%%%%%%%%%%%%%%%%%%%%%%%%%%%%%%%%%%%%%%%%%%%%%%%
%%%%%%%%%%%%%%%%%%%%%%%%%%%%%%%%%%%%%%%%%%%%%%%%%%%%%%
In the preceding section, we discussed various realizations of mini-charged neutrinos within the context of the SM gauge symmetry. The method we have employed to construct these models can also be applied to setups beyond the SM gauge symmetry. To illustrate this, we discuss a realization of mini-charged neutrinos within the context of a left-right symmetric model.

The model is based on the gauge symmetry $SU(3)_C\times SU(2)_L \times SU(2)_R \times U(1)_X$ which also possesses a gaugable global symmetry $U(1)_{B-L}$. Under $\{SU(3)_C,SU(2)_L,SU(2)_R,U(1)_X,U(1)_{B-L}\}$ symmetries, the SM fermion fields transform as follows
\begin{align}
&\ell_{L (R)} = \begin{pmatrix} \nu  \\ e  \end{pmatrix}_{L (R)} \sim(1,2    (1),1 (2),-1,-1), \nonumber\\
&Q_{L (R)} = \begin{pmatrix} u  \\ d  \end{pmatrix}_{L (R)} \sim(3,2 (1),1 (2),\frac{1}{3},\frac{1}{3}).
\end{align}
The masses of these fermions are generated via a generalized see-saw mechanism, for which the model also includes the following vector-like fermions  
\begin{align}
& N \sim(1,1,1,0,-1), \quad E \sim(1,1,1,-2,-1), \nonumber \\
&U \sim(3,1,1,\frac{4}{3},\frac{1}{3}), \quad D \sim(3,1,1,-\frac{2}{3},\frac{1}{3}).
\end{align}
The scalar sector of the model contains two multiplets, which are
\begin{align}
&H_{L (R)} = \begin{pmatrix} H^+  \\ H^0  \end{pmatrix}_{L (R)} \sim(1,2 (1),1 (2),1,0),
\end{align}
where the neutral states acquire a vacuum expectation value $\langle H_{L,R}^0\rangle=v_{L,R}$. It is important to note that, both these scalars transform trivially under the $U(1)_{B-L}$ symmetry, hence after the spontaneous symmetry breaking, the $B-L$ symmetry remains unbroken.  Similar to the previous scenarios, one can gauge a linear combination $X$ and $B-L$, which yields 
\begin{align}
    Q=Q_{st} + \epsilon (B-L),
\end{align}
where $Q_{st}=I_{3L} + I_{3R} + \frac{X}{2}$.

The masses of the fermions are generated via the following Lagrangian terms 
\begin{align}
\label{eq:LaYuk:LR}
    \mathcal{L}_Y =& ~~Y_u\ (\overline{Q}_L \tilde{H}_L + \overline{Q}_R \tilde{H}_R) U + Y_d\ (\overline{Q}_L H_L + \overline{Q}_R H_R) D \nonumber\\
    &+ Y_\ell\ (\overline{\ell}_L H_L + \overline{\ell}_R H_R) E +  Y_{\nu}\ (\overline{\ell}_L \tilde{H}_L+ \overline{\ell}_R \tilde{H}_R) N \nonumber \\
    &+ M_U \overline{U}_L U_R + M_D \overline{D}_L D_R + M_E \overline{E}_L E_R + M_N \overline{N}_L N_R \nonumber \\
    & + h.c,
\end{align}
where the parity symmetry conditions $f_{L} \leftrightarrow f_{R}$ (where $f\equiv\{Q,\ell \}$),  $F_L \leftrightarrow F_R$ (where $F\equiv\{U,D,E,N\}$), and $H_L \leftrightarrow H_R$ are imposed. Eq. \eqref{eq:LaYuk:LR}  yields $6\times 6$ mass matrices for fermions, which are explicitly shown below
\begin{align}
    {\cal M}_{f,F} = 
    \begin{pmatrix}
    0 & Y_f \,\kappa_L \\
    Y_f^\dagger\, \kappa_R & M_{F}
    \end{pmatrix} \,.
    \label{eq:massmatrix}
\end{align}
The corresponding eigenvalues can be obtained by block-diagonalizing $\cal M$ by performing a bi-unitary transformation, leading order which leads
\begin{align}
    &\hat{m_f} \simeq - v_L v_R Y_f M_F^{-1} Y_f^\dagger  \, , \nonumber \\ &\hat{M_F} \simeq M_F + \frac{1}{2}\left(v_R^2 Y_f^{\dagger}Y_f M_F^{-1} + v_L^2 Y_fY_f^{\dagger} M_F^{-1}\right),
    \label{eq:fermionmass}
\end{align}
where $\hat{m_f}$ ($\hat{M_f}$ ) represents mass matrix of light (heavy) fermions states. 

\begin{figure}[thb!]
\includegraphics[width=0.45\textwidth]{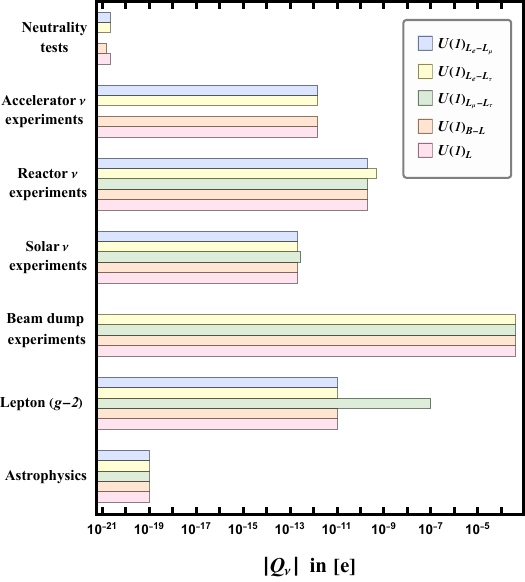}
\caption{Summary of constraints on charged neutrinos. Here blue, yellow, green, orange and purple bars correspond to models based on  $U(1)_{L_e-L_\mu}, U(1)_{L_eL_\tau}, U(1)_{L_\mu-L_\tau}, U(1)_{B-L},$ and $U(1)_L$ gaugable symmetries, respectively.  See text and Tab.~\ref{Tab:Current_Status} for more details.      
}\label{Fig:Current_Status}
\end{figure}

%%%%%%%%%%%%%%%%%%%%%%%%%%%%%%%%%%%%%%%%%%%%%%%%%%%%%%
%%%%%%%%%%%%%%%%%%%%%%%%%%%%%%%%%%%%%%%%%%%%%%%%%%%%%%
\textbf{\emph{Current status of charged neutrinos}--} 
%%%%%%%%%%%%%%%%%%%%%%%%%%%%%%%%%%%%%%%%%%%%%%%%%%%%%%
%%%%%%%%%%%%%%%%%%%%%%%%%%%%%%%%%%%%%%%%%%%%%%%%%%%%%%
Here we summarize the current status of charged neutrinos from various experimental considerations as well as from astrophysics. We primarily focus on five scenarios of charged neutrinos, which are based on $U(1)_{L_e-L_\mu},U(1)_{L_e-L_\tau},U(1)_{L_\mu-L_\tau},U(1)_{B-L},$ and $U(1)_L$ gaugable symmetries, respectively (see Tab.~\ref{tab:gauge_symmetry} for the corresponding charges of matter fields of each of the scenarios). Although the three models defined by $U(1)_{L_i-L_j}$ symmetry are incompatible with the neutrino oscillation data, we include them here for completeness. The results are summarized in Tab.~\ref{Tab:Current_Status} and  Fig.~\ref{Fig:Current_Status}, see the discussion below for details.

%%%%%%%%%%%%%%%%%%%%%%%%%%%%%%%%%%%%%%%%%%%%%

%%%%%%%%%%%%%%%%%%%%%%%%%%%%%%%%%%%%%%%%%%%%%

%%%%%%%%%%%%%%%%%%%%%%%%%%%%%%%%%%%%%%%%%%%%%%%%%%%%%%%%%%%%%%%%%%%%%%%%%%%%%%%%%%%%%%%%%%

%%%%%%%%%%%%%%%%%%%%%%%%%%%%%%%%%%%%%%%%%%%%%%%%%%%%%%%%%%%%%%%%%%%%%%%%%%%%%%%%%%%%%%%%%%
\begin{table*}[th]
\centering
\renewcommand{\arraystretch}{1.9}
\resizebox{1.0\textwidth}{!}{
\begin{tabular}{|c|l|>{\columncolor{blue!10}}c|>{\columncolor{yellow!10}}c|>{\columncolor{green!10}}c|>{\columncolor{orange!10}}c|>{\columncolor{purple!10}}c|}
\hline \hline
 & \cellcolor{lightgray!30} & \multicolumn{5}{c|}{\cellcolor{cyan!30}\textbf{Charge of neutrino in \( [e] \)}} \\ \cline{3-7}
 & \cellcolor{lightgray!30}\textbf{Experiment/Method} & \( \mathbf{U(1)_{L_{e}-L_{\mu}}} \) & \( \mathbf{U(1)_{L_{e}-L_{\tau}}} \) & \( \mathbf{U(1)_{L_{\mu}-L_{\tau}}} \) & \( \mathbf{U(1)_{B-L}} \) & \( \mathbf{U(1)_{L}} \) \\ \hline

% Neutrality test
\multirow{2}{*}{\rotatebox{90}{\parbox[c][1.2cm][c]{1.4cm}{\centering \footnotesize Neutrality\\test}}}
%\vertcell{2}{orange!10}{\parbox[c][1.1cm][c]{1.05cm}{\footnotesize Neutrality\\test}}
%\multirow{2}{*}{\cellcolor{orange!10}\rotatebox{90}{\parbox[c][1.2cm][c]{1.4cm}{\centering \footnotesize Neutrality\\test}}}
& \cellcolor{orange!20}Neutron & --- & --- & --- & \( =(0.4\pm 1.1)\times 10^{-21} \) & --- \\ \cline{2-7}
& \cellcolor{orange!20}Matter & \( =(-0.2\pm 2.3)\times 10^{-21} \) & \( =(-0.2\pm 2.3)\times 10^{-21} \) & --- & \( =(0.2\pm 2.1)\times 10^{-21} \) & \( =(-0.2\pm 2.3)\times 10^{-21} \) \\ \hline

% Accelerator ν experiment
%\vertcell{5}{pink!10}{\footnotesize Accelerator \(\nu\) \\ experiment}
\multirow{5}{*}{\rotatebox{90}{{\footnotesize \shortstack{Accelerator \(\nu\) \\ experiment}}}} 
& \cellcolor{pink!10}TEXONO (2002) & \( < 3.7\times 10^{-12} \) & \( < 3.7\times 10^{-12} \) & --- & \( < 3.7\times 10^{-12} \) & \( < 3.7\times 10^{-12} \) \\ \cline{2-7}
& \cellcolor{pink!10}GEMMA & \( < 1.5\times 10^{-12} \) & \( < 1.5\times 10^{-12} \) & --- & \( < 1.5\times 10^{-12} \) & \( < 1.5\times 10^{-12} \) \\ \cline{2-7}
& \cellcolor{pink!10}TEXONO (2014) & \( < 2.1\times 10^{-12} \) & \( < 2.1\times 10^{-12} \) & --- & \( < 2.1\times 10^{-12} \) & \( < 2.1\times 10^{-12} \) \\ \cline{2-7}
& \cellcolor{pink!10}CONUS & \( < 3.3\times 10^{-12} \) & \( < 3.3\times 10^{-12} \) & --- & \( < 3.3\times 10^{-12} \) & \( < 3.3\times 10^{-12} \) \\ \cline{2-7}
& \cellcolor{pink!10}Dresden-II & \( \in (-9.3,9.5)\times 10^{-12} \) & \( \in (-9.3,9.5)\times 10^{-12} \) & --- & \( \in (-9.3,9.5)\times 10^{-12} \) & \( \in (-9.3,9.5)\times 10^{-12} \) \\ \hline

% Reactor ν experiment
\multirow{3}{*}{\rotatebox{90}{{\footnotesize \shortstack{Reactor \(\nu\) \\ experiment}}}} 
&\cellcolor{yellow!20} LSND & \( < 3\times 10^{-9} \) & --- & \( < 3\times 10^{-9} \) & \( < 3\times 10^{-9} \) & \( < 3\times 10^{-9} \) \\ \cline{2-7}
& \cellcolor{yellow!20} DONUT & --- & \( < 4\times 10^{-6} \) & \( < 4\times 10^{-6} \) & \( < 4\times 10^{-6} \) & \( < 4\times 10^{-6} \) \\ \cline{2-7}
& \cellcolor{yellow!20}COHERENT & \( \in (-1.9,1.9)\times 10^{-10} \) & \( \in (-5.0,5.0)\times 10^{-10} \) & \( \in (-1.9,1.9)\times 10^{-10} \) & \( \in (-1.9,1.9)\times 10^{-10} \) & \( \in (-1.9,1.9)\times 10^{-10} \) \\ \hline

% Solar ν experiment
\multirow{4}{*}{\rotatebox{90}{{\footnotesize \shortstack{Solar \(\nu\) \\ experiment}}}} 
& \cellcolor{magenta!10}XMAS-I & \( < 7.3\times 10^{-12} \) & \( < 7.3\times 10^{-12} \) & \( < 1.1\times 10^{-11} \) & \( < 5.4\times 10^{-12} \) & \( < 5.4\times 10^{-12} \) \\ \cline{2-7}
& \cellcolor{magenta!10}LUX-ZEPLIN & \( \in (-2.1,2.0)\times 10^{-13} \) & \( \in (-2.1,2.0)\times 10^{-13} \) & \( \in (-2.8,2.8)\times 10^{-13} \) & \( \in (-2.1,2.0)\times 10^{-13} \) & \( \in (-2.1,2.0)\times 10^{-13} \) \\ \cline{2-7}
& \cellcolor{magenta!10}XENONnT & \( \in (-6.2,6.1)\times 10^{-13} \) & \( \in (-5.4,5.2)\times 10^{-13} \) & \( \in (-5.4,5.2)\times 10^{-13} \) & \( \in (-5.4,5.2)\times 10^{-13} \) & \( \in (-5.4,5.2)\times 10^{-13} \) \\ \cline{2-7}
& \cellcolor{magenta!10}PandaX-4T & \( \in (-1.3,1.6)\times 10^{-12} \) & \( \in (-1.3,1.6)\times 10^{-12} \) & \( \in (-2.2,2.2)\times 10^{-12} \) & \( \in (-1.3,1.6)\times 10^{-12} \) & \( \in (-1.3,1.6)\times 10^{-12} \) \\ \hline

% Beam Dump
\rotatebox{90}{\parbox[c][1.2cm][c]{0.9cm}{\centering \footnotesize \shortstack{Beam\\Dump}}}
& \cellcolor{olive!10}BEBC & --- & \( < 4\times 10^{-4} \) & \( < 4\times 10^{-4} \) & \( < 4\times 10^{-4} \) & \( < 4\times 10^{-4} \) \\ \hline

% g-2
\multirow{2}{*}{\rotatebox{90}{{\footnotesize \shortstack{\((g-2)_\ell\) }}}} 
& \cellcolor{teal!10}Muon \( (g-2) \) & \( < 10^{-7} \) & --- & \( < 10^{-7} \) & \( < 10^{-7} \) & \( < 10^{-7} \) \\ \cline{2-7}
& \cellcolor{teal!10}Electron \( (g-2) \) & \( < 10^{-11} \) & \( < 10^{-11} \) & --- & \( < 10^{-11} \) & \( < 10^{-11} \) \\ \hline

% Astrophysics
\multirow{5}{*}{\rotatebox{90}{{\footnotesize Astrophysics}}} 
& \cellcolor{brown!10}SN1987A & \( \lesssim 10^{-17}-10^{-15} \) & \( \lesssim 10^{-17}-10^{-15} \) & --- & \( \lesssim 10^{-17}-10^{-15} \) & \( \lesssim 10^{-17}-10^{-15} \) \\ \cline{2-7}
& \cellcolor{brown!10}Solar cooling & \( \lesssim 4\times 10^{-14} \) & \( \lesssim 4\times 10^{-14} \) & \( \lesssim 4\times 10^{-14} \) & \( \lesssim 3\times 10^{-14} \) & \( \lesssim 3\times 10^{-14} \) \\ \cline{2-7}
& \cellcolor{brown!10}TRGB & \( <6.3\times 10^{-15} \) & \( <6.3\times 10^{-15} \) & \( <6.3\times 10^{-15} \) & \( <6.3\times 10^{-15} \) & \( <6.3\times 10^{-15} \) \\ \cline{2-7}
& \cellcolor{brown!10}Magnetars & \( <10^{-12}-10^{-11} \) & \( <10^{-12}-10^{-11} \) & \( <10^{-12}-10^{-11} \) & \( <10^{-12}-10^{-11} \) & \( <10^{-12}-10^{-11} \) \\ \cline{2-7}
& \cellcolor{brown!10}Pulsars & \( < 10^{-19} \) & \( < 10^{-19} \) & \( < 10^{-19} \) & \( < 10^{-19} \) & \( < 10^{-19} \) \\ \hline \hline
\end{tabular}}
\caption{The most up-to-date constraints on neutrino electric charges arise from laboratory experiments, including neutrino experiments, beam dump searches, neutrality tests, lepton $(g-2)$ measurements, as well as from astrophysical observations. See text for more details.}
\label{Tab:Current_Status}
\end{table*}

Neutrino scattering experiments can directly probe the electric charge of neutrinos. The observation of no significant deviation from the canonical SM prediction from these experiments imposes strong limits on $Q_{\nu}$. These bounds are generally flavor-dependent, meaning they differ for different flavors of neutrinos depending on the experimental searches.
%These bounds may vary for different flavors of neutrinos depending on the experimental searches.
For instance, reactor based experiments provide bounds on $Q_{\nu_e}$  by studying the elastic scattering process of $\bar{\nu_e}$ with electrons/nuclei. The corresponding limits from TEXONO~\cite{TEXONO:2002pra} and Dresden-II~\cite{Colaresi:2022obx} are given below:
\begin{align}
&|Q_{\nu_e}|<1.0\times 10^{-12}e & \mathrm{TEXONO}\, \mbox{\cite{Chen:2014dsa}}, \\
& Q_{\nu_e}\in (-9.3,9.5)\times 10^{-12} e & \mathrm{Dresden\text-II}\, \mbox{\cite{AtzoriCorona:2022qrf}}. 
\end{align}
On the other hand, accelerator based experiment COHERENT~\cite{COHERENT:2020iec,COHERENT:2021xmm} provide bounds on  $Q_{\nu_{e,\mu}}$~\cite{AtzoriCorona:2022qrf}:
\begin{align}
 &Q_{\nu_e}\in (-5.0,5.0)\times 10^{-10}e, \\&Q_{\nu_{\mu}}\in (-1.9,1.9)\times 10^{-10}e.  
\end{align}
Whereas solar neutrino experiments impose constraints on the charges of all flavors of neutrinos:
\begin{align}
    \mathrm{XMAS\text-I} \, \mbox{\cite{XMASS:2020zke}} &
    \begin{cases}
            |Q_{\nu_e}|<7.3\times 10^{-12}e, \\
            |Q_{\nu_{\mu,\tau}}|<1.1\times 10^{-12}e,\\
            |Q_{\nu}|<5.4\times 10^{-12}e,
    \end{cases}\\
    \mathrm{LUX\text-ZEPLIN}\, \mbox{\cite{LZ:2022lsv,AtzoriCorona:2022jeb}} &
    \begin{cases}
            Q_{\nu_e}\in (-2.1,2.0)\times 10^{-13}e, \\
            |Q_{\nu_{\mu}}|<3.1\times 10^{-13}e,\\
            |Q_{\nu_{\tau}}|<2.8\times 10^{-13}e,
    \end{cases}\\
    \mathrm{XENONnT}\, \mbox{\cite{XENON:2022ltv,A:2022acy}} &
    \begin{cases}
            Q_{\nu_e}\in (-1.3,6.4)\times 10^{-13}e, \\
            Q_{\nu_{\mu}}\in (-6.2,6.1)\times 10^{-13}e,\\
            Q_{\nu_{\tau}}\in (-5.4,5.2)\times 10^{-13}e,
    \end{cases}\\
    \mathrm{PandaX\text-4T}\, \mbox{\cite{PandaX:2022ood,Giunti:2023yha}} &
    \begin{cases}
            Q_{\nu_e}\in (-1.3,1.6)\times 10^{-12}e, \\
            Q_{\nu_{\mu,\tau}}\in (-2.2,2.2)\times 10^{-12}e.
    \end{cases}
\end{align}
We recast these model-independent limits for our benchmark scenarios, see Tab.~\ref{Tab:Current_Status}. For this, we have taken the most stringent bound applicable to the concerned scenario from each of the experiments. This approach is justified since in all of the benchmark scenarios that we are focused on, the electric charge of the neutrinos depends only on a single parameter $\epsilon$, see Tab.~\ref{tab:gauge_symmetry}. %Further limits on $Q_{\nu}$ can be derived by utilizing the bounds on neutrino magnetic moment from various low-energy neutrino scattering experiments.
Additional constraints from neutrino scattering experiments can be derived by reinterpreting the existing bounds imposed by these experiments on the neutrino magnetic moment in terms of 
$Q_{\nu}$~\cite{Babu:1993yh,Gninenko:2006fi,Studenikin:2013my}.
%In addition to these constraints, one can utilize the bounds on neutrino magnetic moment from various low-energy neutrino scattering experiments to derive limits on $Q_{\nu}$~\cite{Babu:1993yh,Gninenko:2006fi,Studenikin:2013my}. 
%This is achieved by comparing the cross sections of neutrino scattering process due to the magnetic moment of neutrino with the corresponding contribution induced by the electric charge of neutrino.  
This yields several strong limits, which are given below:
\begin{align}
 & |Q_{\nu_\tau}|<4\times 10^{-4} e &\mathrm{BEBC}~\mbox{\cite{BEBCWA66:1986err,Cooper-Sarkar:1991vsl,Babu:1993yh}}, \\
& |Q_{\nu_e}|<3.7\times 10^{-12} e &\mathrm{TEXONO}~\mbox{\cite{TEXONO:2002pra,Gninenko:2006fi}} \\
& |Q_{\nu_e}|<1.5\times 10^{-12} e &\mathrm{GEMMA}~\mbox{\cite{Beda:2012zz,Studenikin:2013my}}, 
\\
& |Q_{\nu_e}|<3.3\times 10^{-12} e &\mathrm{CONUS}~\mbox{\cite{CONUS:2022qbb}},
\\
& |Q_{\nu_\mu}|<3\times 10^{-9} e &\mathrm{LSND}~\mbox{\cite{LSND:2001akn,Das:2020egb}},
\\
& |Q_{\nu_\tau}|<4\times 10^{-6} e &\mathrm{DONUT}~\mbox{\cite{DONUT:2001zvi,Das:2020egb}}.
\end{align}
See Tab.~\ref{Tab:Current_Status} for the corresponding bounds for the benchmark scenarios.

The most stringent bounds come from the neutrality tests of neutrons and matter.
Although these measurements have no direct implications for $Q_{\nu}$, they yield strong limits on models of charged neutrinos. This is a direct consequence of the modification of charges of matter fields in neutrino charge models, see Tab.~\ref{tab:gauge_symmetry}. Except for the scenario of $U(1)_{L_{\mu}-L_{\tau}}$ symmetry, in all other cases, the charges of the SM first-generation particles are altered from their canonical values. This leads to deviations of neutron charge (denoted as $Q_n$) and matter charge (denoted as $Q_m$) from neutrality. 
%This leads to deviations from the charge neutrality of matter and the neutron. 
Moreover, such deviations are also directly related to $Q_{\nu}$ in these models. Hence, the bounds on $Q_{m,n}$  also yield constraints on
$Q_{\nu}$ for these scenarios. To see this explicitly, consider the $U(1)_{B-L}$ model, wherein  $Q_n=-Q_{\nu}$. Therefore, the precise measurement of  neutron charge $Q_n=(-0.4\pm 1.1)\times 10^{-21} e$ \cite{PhysRevD.37.3107} yields $Q_{\nu}=(0.4\pm 1.1)\times 10^{-21} e$ for this scenario. In contrast, $Q_n=0$ for the other three models; hence, this measurement imposes no constraint for these scenarios. On the other hand, neutrality tests of matter provide limits for all four models. For these scenarios, one can write $Q_m$ in terms of $Q_{\nu}$ as follows:
\begin{align}
\label{eq:Qm}
    Q_m &  =   
    \begin{cases}
        \frac{Z Q_{\nu}}{A}  & U(1)_X\equiv U(1)_{L_{e}-L_{\mu}}, U(1)_{L_{e}-L_{\tau}}, U(1)_L, \\
        \frac{-N Q_{\nu}}{A}  & U(1)_X\equiv U(1)_{B-L}.
    \end{cases}
\end{align}
Here we have taken $Q_m\equiv\frac{Z\left(Q_p+Q_e\right)+ N Q_n }{A}$, where $A$,  $Z$, and  $N$ denote atomic mass, atomic number, and neutron number of the matter, respectively. Here $Q_p$ stands for the charge of the proton. Ref.~\cite{PhysRevA.83.052101} provides most stringent bound on $Q_m=(-0.1\pm 1.1)\times 10^{-21} e$. This can be translated for the benchmark models by using Eq.~\eqref{eq:Qm},  see Tab.~\ref{Tab:Current_Status}.

In addition to this, non-standard charges of matter fields also have an impact on their anomalous magnetic moment contributions.
Hence, the precise measurements of  muon and electron $(g-2)$  indirectly set constraints on scenarios of charged neutrinos. To estimate this, we calculate the leading order deviation of lepton anomalous magnetic moment ($\delta a_{\ell}$) from their canonical value, giving~\cite{Babu:1992sw}  
\begin{align}
    |\delta a_{\ell}|\simeq \frac{3|\epsilon| \alpha_{\mathrm{em}}}{2\pi}\frac{e}{2m_{\ell}}.
\end{align}
Then, we compare this with the error margin in measurement of $a_{\ell}$ ($\delta a^{\mathrm{exp}}_{\ell}$): $\delta a_{\ell}<\delta a^{\mathrm{exp}}_{\ell}$. Here we assume that the SM prediction of $a_{\ell}$ is in align with the measured value of $a_{\ell}$. 
The resulting limit from muon  $(g-2)$~\cite{Muong-2:2023cdq}  yields $|\epsilon|=|Q_{\nu}|<10^{-7}$ for $U(1)_{L_e-L_{\mu}}, U(1)_{L_{\mu}-L_{\tau}}, U(1)_{B-L}$, and $U(1)_L$ scenarios. Similarly, a limit from electron $(g-2)$~\cite{ParticleDataGroup:2024cfk} can be obtained for $U(1)_{L_e-L_{\mu}},  U(1)_{L_{e}-L_{\tau}}, U(1)_{B-L}$, and $U(1)_L$ scenarios: $|Q_{\nu}|<10^{-11}$.

Strong constraints on $Q_{\nu}$ also arise from various astrophysical considerations. In stellar environments, the tree-level coupling of photons with charged neutrinos induces plasmon decay ($\gamma \xrightarrow{}\nu\bar{\nu}$), which constitutes an additional energy loss for stars. Such non-standard contributions are severely constrained and yield stringent limits on $Q_{\nu}$. The resulting limit from red giant stars has been derived in Ref.~\cite{Fung:2023euv}:  $|Q_{\nu}|<6.3\times 10^{-15}e$. Similarly, a limit from the Sun for a single generation of charged neutrino is analytically derived in Ref.~\cite{Raffelt:1999gv}, which can be readily translated for scenarios of two charged neutrinos ($U(1)_{L_i-L_j}$ models) and three charged neutrinos (both $U(1)_{B-L}$ and $U(1)_L$ models), see Tab.~\ref{Tab:Current_Status}. Limit also arises from SN1987A measurements \cite{Barbiellini:1987zz}. For deriving this bound, they compared the arrival interval of charged neutrinos emitted from the supernovae with the observed time duration $\sim 10$ sec \cite{Kamiokande-II:1987idp}, which yielded a bound on $|Q_{\nu_e}|\lesssim (10^{-17}-10^{-15})e$. An even stronger bound  can be obtained by taking into account the effect of charged neutrinos on the rotation frequency of pulsars : $|Q_{\nu}|< 10^{-19}e$ \cite{Studenikin:2012vi}. An additional constraint  $|Q_{\nu}|< (10^{-12}-10^{-11})e$ comes from the consideration of charged neutrino productions in magnetars through the Schwinger mechanism \cite{Das:2020egb}. We recast all these limits for the benchmark models, see
Tab.~\ref{Tab:Current_Status}.

%%%%%%%%%%%%%%%%%%%%%%%%%%%%%%%%%%%%%%%%%%%%%%%%%%%%%%%%%%%%%%%%%%%%%%%%%%%%%%%%%%%%%%%%%%

In summary, a wide range of ongoing and upcoming direct dark matter detection and neutrino scattering experiments are expected to achieve unprecedented sensitivity in the coming decades. If neutrinos carry a small electric charge, this could lead to observable effects. Experiments like DUNE \cite{Mathur:2021trm} and the LHC Forward Physics Facility \cite{MammenAbraham:2023psg, Ismail:2021dyp} may also offer promising sensitivity to such charges. Additional avenues for probing neutrino charges include studying flux attenuation in future neutrino telescopes \cite{Huang:2021mki, Huang:2022pce} and analyzing supernova neutronization burst signals \cite{Jana:2022tsa, Jana:2023ufy}. These directions, however, lie beyond the scope of the present study and will be pursued in future work.

%%%%%%%%%%%%%%%%%%%%%%%%%%%%%%%%%%%%%%%%%%%%%%%%%%%%%%
%%%%%%%%%%%%%%%%%%%%%%%%%%%%%%%%%%%%%%%%%%%%%%%%%%%%%%
\textbf{\emph{Conclusions}--} 
%%%%%%%%%%%%%%%%%%%%%%%%%%%%%%%%%%%%%%%%%%%%%%%%%%%%%%
%%%%%%%%%%%%%%%%%%%%%%%%%%%%%%%%%%%%%%%%%%%%%%%%%%%%%%
We have systematically examined scenarios of minicharged neutrinos within the Standard Model framework, respecting electromagnetic gauge invariance. Both flavor-universal and non-universal charge assignments were explored, with only the former aligning with neutrino oscillation data. By incorporating various theoretical and experimental constraints, we established upper bounds on neutrino charges: $10^{-21}e$ for models based on $L_e - L_\mu$, $L_e - L_\tau$, and $L$ symmetries (driven by matter neutrality), $10^{-19}e$ for the $L_\mu - L_\tau$ case (constrained by astrophysics), and $10^{-21}e$ for the $B - L$ scenario (from neutron and matter neutrality). This work outlines a roadmap for how neutrinos could carry small electric charges in different theories and highlights the profound implications such charges would have on our understanding of charge quantization.

\vspace{0.1in}
%%%%%%%%%%%%%%%%%%%%%%%%%%%%%%%%%%%%%%%%%%%%%%%%%%%%%%
\begin{acknowledgments}
{\textbf {\textit {Acknowledgments.--}}} SJ thanks Tatsu Takeuchi and Debajyoti Choudhury for insightful discussions. SJ acknowledges the support provided by the Department of Atomic Energy, Government of India, through the Harish-Chandra Research Institute. SJ is also grateful to the CERN Theory Group for their kind hospitality during the final stages of this work. 

\end{acknowledgments}

\bibliographystyle{utphys}
\bibliography{reference}

\providecommand{\href}[2]{#2}\begingroup\raggedright\begin{thebibliography}{10}

\bibitem{Babu:1989tq}
K.~S. Babu and R.~N. Mohapatra, ``{Is There a Connection Between Quantization of Electric Charge and a Majorana Neutrino?},'' \href{http://dx.doi.org/10.1103/PhysRevLett.63.938}{{\em Phys. Rev. Lett.} {\bfseries 63} (1989) 938}.

\bibitem{Foot:1998tb}
R.~Foot and R.~R. Volkas, ``{Do the Super-Kamiokande atmospheric neutrino results explain electric charge quantization?},'' \href{http://dx.doi.org/10.1103/PhysRevD.59.097301}{{\em Phys. Rev. D} {\bfseries 59} (1999) 097301}, \href{http://arxiv.org/abs/hep-ph/9808388}{{\ttfamily arXiv:hep-ph/9808388}}.

\bibitem{Ignatiev:1994fd}
A.~Y. Ignatiev and G.~C. Joshi, ``{The Charged neutrino: A New approach to the solar neutrino problem},'' \href{http://dx.doi.org/10.1142/S0217732394001313}{{\em Mod. Phys. Lett. A} {\bfseries 9} (1994) 1479--1488}, \href{http://arxiv.org/abs/hep-ph/9403332}{{\ttfamily arXiv:hep-ph/9403332}}.

\bibitem{Wolfenstein:1977ue}
L.~Wolfenstein, ``{Neutrino Oscillations in Matter},'' \href{http://dx.doi.org/10.1103/PhysRevD.17.2369}{{\em Phys. Rev. D} {\bfseries 17} (1978) 2369--2374}.

\bibitem{Mikheev:1986wj}
S.~P. Mikheev and A.~Y. Smirnov, ``{Resonant amplification of neutrino oscillations in matter and solar neutrino spectroscopy},'' \href{http://dx.doi.org/10.1007/BF02508049}{{\em Nuovo Cim. C} {\bfseries 9} (1986) 17--26}.

\bibitem{Mikheyev:1985zog}
S.~P. Mikheyev and A.~Y. Smirnov, ``{Resonance Amplification of Oscillations in Matter and Spectroscopy of Solar Neutrinos},'' {\em Sov. J. Nucl. Phys.} {\bfseries 42} (1985) 913--917.

\bibitem{Maltoni:2015kca}
M.~Maltoni and A.~Y. Smirnov, ``{Solar neutrinos and neutrino physics},'' \href{http://dx.doi.org/10.1140/epja/i2016-16087-0}{{\em Eur. Phys. J. A} {\bfseries 52} no.~4, (2016) 87}, \href{http://arxiv.org/abs/1507.05287}{{\ttfamily arXiv:1507.05287 [hep-ph]}}.

\bibitem{Giunti:2014ixa}
C.~Giunti and A.~Studenikin, ``{Neutrino electromagnetic interactions: a window to new physics},'' \href{http://dx.doi.org/10.1103/RevModPhys.87.531}{{\em Rev. Mod. Phys.} {\bfseries 87} (2015) 531}, \href{http://arxiv.org/abs/1403.6344}{{\ttfamily arXiv:1403.6344 [hep-ph]}}.

\bibitem{Babu:2020ivd}
K.~S. Babu, S.~Jana, and M.~Lindner, ``{Large Neutrino Magnetic Moments in the Light of Recent Experiments},'' \href{http://dx.doi.org/10.1007/JHEP10(2020)040}{{\em JHEP} {\bfseries 10} (2020) 040}, \href{http://arxiv.org/abs/2007.04291}{{\ttfamily arXiv:2007.04291 [hep-ph]}}.

\bibitem{Babu:2021jnu}
K.~S. Babu, S.~Jana, M.~Lindner, and V.~P. K, ``{Muon g \ensuremath{-} 2 anomaly and neutrino magnetic moments},'' \href{http://dx.doi.org/10.1007/JHEP10(2021)240}{{\em JHEP} {\bfseries 10} (2021) 240}, \href{http://arxiv.org/abs/2104.03291}{{\ttfamily arXiv:2104.03291 [hep-ph]}}.

\bibitem{Bernabeu:2000hf}
J.~Bernabeu, L.~G. Cabral-Rosetti, J.~Papavassiliou, and J.~Vidal, ``{On the charge radius of the neutrino},'' \href{http://dx.doi.org/10.1103/PhysRevD.62.113012}{{\em Phys. Rev. D} {\bfseries 62} (2000) 113012}, \href{http://arxiv.org/abs/hep-ph/0008114}{{\ttfamily arXiv:hep-ph/0008114}}.

\bibitem{Foot:1990mn}
R.~Foot, ``{New Physics From Electric Charge Quantization?},'' \href{http://dx.doi.org/10.1142/S0217732391000543}{{\em Mod. Phys. Lett. A} {\bfseries 6} (1991) 527--530}.

\bibitem{Babu:1992sw}
K.~S. Babu and R.~R. Volkas, ``{Bounds on minicharged neutrinos in the minimal Standard Model},'' \href{http://dx.doi.org/10.1103/PhysRevD.46.R2764}{{\em Phys. Rev. D} {\bfseries 46} (1992) R2764--R2768}, \href{http://arxiv.org/abs/hep-ph/9208260}{{\ttfamily arXiv:hep-ph/9208260}}.

\bibitem{Giunti:2024gec}
C.~Giunti, K.~Kouzakov, Y.-F. Li, and A.~Studenikin, ``{Neutrino Electromagnetic Properties},'' \href{http://arxiv.org/abs/2411.03122}{{\ttfamily arXiv:2411.03122 [hep-ph]}}.

\bibitem{Lee:1956qn}
T.~D. Lee and C.-N. Yang, ``{Question of Parity Conservation in Weak Interactions},'' \href{http://dx.doi.org/10.1103/PhysRev.104.254}{{\em Phys. Rev.} {\bfseries 104} (1956) 254--258}.

\bibitem{Foot:1991py}
R.~Foot, H.~Lew, and R.~R. Volkas, ``{Possible consequences of parity conservation},'' \href{http://dx.doi.org/10.1142/S0217732392004031}{{\em Mod. Phys. Lett. A} {\bfseries 7} (1992) 2567--2574}.

\bibitem{Berezhiani:1995yi}
Z.~G. Berezhiani and R.~N. Mohapatra, ``{Reconciling present neutrino puzzles: Sterile neutrinos as mirror neutrinos},'' \href{http://dx.doi.org/10.1103/PhysRevD.52.6607}{{\em Phys. Rev. D} {\bfseries 52} (1995) 6607--6611}, \href{http://arxiv.org/abs/hep-ph/9505385}{{\ttfamily arXiv:hep-ph/9505385}}.

\bibitem{Silagadze:1995tr}
Z.~K. Silagadze, ``{Neutrino mass and the mirror universe},'' {\em Phys. Atom. Nucl.} {\bfseries 60} (1997) 272--275, \href{http://arxiv.org/abs/hep-ph/9503481}{{\ttfamily arXiv:hep-ph/9503481}}.

\bibitem{Farzan:2012sa}
Y.~Farzan and E.~Ma, ``{Dirac neutrino mass generation from dark matter},'' \href{http://dx.doi.org/10.1103/PhysRevD.86.033007}{{\em Phys. Rev. D} {\bfseries 86} (2012) 033007}, \href{http://arxiv.org/abs/1204.4890}{{\ttfamily arXiv:1204.4890 [hep-ph]}}.

\bibitem{Ma:2014qra}
E.~Ma and R.~Srivastava, ``{Dirac or inverse seesaw neutrino masses with $B-L$ gauge symmetry and $S_3$ flavor symmetry},'' \href{http://dx.doi.org/10.1016/j.physletb.2014.12.049}{{\em Phys. Lett. B} {\bfseries 741} (2015) 217--222}, \href{http://arxiv.org/abs/1411.5042}{{\ttfamily arXiv:1411.5042 [hep-ph]}}.

\bibitem{Ma:2015mjd}
E.~Ma, N.~Pollard, R.~Srivastava, and M.~Zakeri, ``{Gauge $B-L$ Model with Residual $Z_3$ Symmetry},'' \href{http://dx.doi.org/10.1016/j.physletb.2015.09.010}{{\em Phys. Lett. B} {\bfseries 750} (2015) 135--138}, \href{http://arxiv.org/abs/1507.03943}{{\ttfamily arXiv:1507.03943 [hep-ph]}}.

\bibitem{CentellesChulia:2016rms}
S.~Centelles~Chuli\'a, E.~Ma, R.~Srivastava, and J.~W.~F. Valle, ``{Dirac Neutrinos and Dark Matter Stability from Lepton Quarticity},'' \href{http://dx.doi.org/10.1016/j.physletb.2017.01.070}{{\em Phys. Lett. B} {\bfseries 767} (2017) 209--213}, \href{http://arxiv.org/abs/1606.04543}{{\ttfamily arXiv:1606.04543 [hep-ph]}}.

\bibitem{Bonilla:2016diq}
C.~Bonilla, E.~Ma, E.~Peinado, and J.~W.~F. Valle, ``{Two-loop Dirac neutrino mass and WIMP dark matter},'' \href{http://dx.doi.org/10.1016/j.physletb.2016.09.027}{{\em Phys. Lett. B} {\bfseries 762} (2016) 214--218}, \href{http://arxiv.org/abs/1607.03931}{{\ttfamily arXiv:1607.03931 [hep-ph]}}.

\bibitem{Ma:2016mwh}
E.~Ma and O.~Popov, ``{Pathways to Naturally Small Dirac Neutrino Masses},'' \href{http://dx.doi.org/10.1016/j.physletb.2016.11.027}{{\em Phys. Lett. B} {\bfseries 764} (2017) 142--144}, \href{http://arxiv.org/abs/1609.02538}{{\ttfamily arXiv:1609.02538 [hep-ph]}}.

\bibitem{Wang:2016lve}
W.~Wang and Z.-L. Han, ``{Naturally Small Dirac Neutrino Mass with Intermediate $SU(2)_{L}$ Multiplet Fields},'' \href{http://dx.doi.org/10.1007/JHEP04(2017)166}{{\em JHEP} {\bfseries 04} (2017) 166}, \href{http://arxiv.org/abs/1611.03240}{{\ttfamily arXiv:1611.03240 [hep-ph]}}.

\bibitem{Borah:2017dmk}
D.~Borah and B.~Karmakar, ``{$A_4$ flavour model for Dirac neutrinos: Type I and inverse seesaw},'' \href{http://dx.doi.org/10.1016/j.physletb.2018.03.047}{{\em Phys. Lett. B} {\bfseries 780} (2018) 461--470}, \href{http://arxiv.org/abs/1712.06407}{{\ttfamily arXiv:1712.06407 [hep-ph]}}.

\bibitem{Jana:2019mez}
S.~Jana, P.~K. Vishnu, and S.~Saad, ``{Minimal dirac neutrino mass models from $\hbox {U}(1)_{\mathrm{R}}$ gauge symmetry and left\textendash{}right asymmetry at colliders},'' \href{http://dx.doi.org/10.1140/epjc/s10052-019-7441-9}{{\em Eur. Phys. J. C} {\bfseries 79} no.~11, (2019) 916}, \href{http://arxiv.org/abs/1904.07407}{{\ttfamily arXiv:1904.07407 [hep-ph]}}.

\bibitem{Gu:2019ogb}
P.-H. Gu, ``{Double type II seesaw mechanism accompanied by Dirac fermionic dark matter},'' \href{http://dx.doi.org/10.1103/PhysRevD.101.015006}{{\em Phys. Rev. D} {\bfseries 101} no.~1, (2020) 015006}, \href{http://arxiv.org/abs/1907.10019}{{\ttfamily arXiv:1907.10019 [hep-ph]}}.

\bibitem{Jana:2019mgj}
S.~Jana, P.~K. Vishnu, and S.~Saad, ``{Minimal realizations of Dirac neutrino mass from generic one-loop and two-loop topologies at $d = 5$},'' \href{http://dx.doi.org/10.1088/1475-7516/2020/04/018}{{\em JCAP} {\bfseries 04} (2020) 018}, \href{http://arxiv.org/abs/1910.09537}{{\ttfamily arXiv:1910.09537 [hep-ph]}}.

\bibitem{Wolfenstein:1981kw}
L.~Wolfenstein, ``{Different Varieties of Massive Dirac Neutrinos},'' \href{http://dx.doi.org/10.1016/0550-3213(81)90096-1}{{\em Nucl. Phys. B} {\bfseries 186} (1981) 147--152}.

\bibitem{Petcov:1982ya}
S.~T. Petcov, ``{On Pseudodirac Neutrinos, Neutrino Oscillations and Neutrinoless Double beta Decay},'' \href{http://dx.doi.org/10.1016/0370-2693(82)91246-1}{{\em Phys. Lett. B} {\bfseries 110} (1982) 245--249}.

\bibitem{Valle:1983dk}
J.~W.~F. Valle and M.~Singer, ``{Lepton Number Violation With Quasi Dirac Neutrinos},'' \href{http://dx.doi.org/10.1103/PhysRevD.28.540}{{\em Phys. Rev. D} {\bfseries 28} (1983) 540}.

\bibitem{Kobayashi:2000md}
M.~Kobayashi and C.~S. Lim, ``{Pseudo Dirac scenario for neutrino oscillations},'' \href{http://dx.doi.org/10.1103/PhysRevD.64.013003}{{\em Phys. Rev. D} {\bfseries 64} (2001) 013003}, \href{http://arxiv.org/abs/hep-ph/0012266}{{\ttfamily arXiv:hep-ph/0012266}}.

\bibitem{Babu:2022ikf}
K.~S. Babu, X.-G. He, M.~Su, and A.~Thapa, ``{Naturally light Dirac and pseudo-Dirac neutrinos from left-right symmetry},'' \href{http://dx.doi.org/10.1007/JHEP08(2022)140}{{\em JHEP} {\bfseries 08} (2022) 140}, \href{http://arxiv.org/abs/2205.09127}{{\ttfamily arXiv:2205.09127 [hep-ph]}}.

\bibitem{Biswas:2024wbz}
S.~Biswas, V.~P. K., and A.~Thapa, ``{Connecting pseudo-Nambu-Goldstone dark matter with pseudo-Dirac neutrinos in a left-right symmetry model},'' \href{http://arxiv.org/abs/2407.05482}{{\ttfamily arXiv:2407.05482 [hep-ph]}}.

\bibitem{Foot:1989fh}
R.~Foot, G.~C. Joshi, H.~Lew, and R.~R. Volkas, ``{CHARGED NEUTRINOS?},'' \href{http://dx.doi.org/10.1142/S0217732390000123}{{\em Mod. Phys. Lett. A} {\bfseries 5} (1990) 95}. [Erratum: Mod.Phys.Lett.A 5, 2085 (1990)].

\bibitem{Foot:1992ui}
R.~Foot, H.~Lew, and R.~R. Volkas, ``{Electric charge quantization},'' \href{http://dx.doi.org/10.1088/0954-3899/19/3/005}{{\em J. Phys. G} {\bfseries 19} (1993) 361--372}, \href{http://arxiv.org/abs/hep-ph/9209259}{{\ttfamily arXiv:hep-ph/9209259}}. [Erratum: J.Phys.G 19, 1067 (1993)].

\bibitem{Note1}
See Refs.~\cite {Maruno:1991vr, Takasugi:1991ai, Takasugi:1991wa, Ignatiev:1996np, Ignatiev:1997pk, Babu:1990vw} for scenarios of charged neutrinos emerging from the breaking of electromagnetic gauge symmetry.

\bibitem{He:1990pn}
X.~G. He, G.~C. Joshi, H.~Lew, and R.~R. Volkas, ``{NEW Z-prime PHENOMENOLOGY},'' \href{http://dx.doi.org/10.1103/PhysRevD.43.R22}{{\em Phys. Rev. D} {\bfseries 43} (1991) 22--24}.

\bibitem{Chao:2010mp}
W.~Chao, ``{Pure Leptonic Gauge Symmetry, Neutrino Masses and Dark Matter},'' \href{http://dx.doi.org/10.1016/j.physletb.2010.10.056}{{\em Phys. Lett. B} {\bfseries 695} (2011) 157--161}, \href{http://arxiv.org/abs/1005.1024}{{\ttfamily arXiv:1005.1024 [hep-ph]}}.

\bibitem{TEXONO:2002pra}
{\bfseries TEXONO} Collaboration, H.~B. Li {\em et~al.}, ``{Limit on the electron neutrino magnetic moment from the Kuo-Sheng reactor neutrino experiment},'' \href{http://dx.doi.org/10.1103/PhysRevLett.90.131802}{{\em Phys. Rev. Lett.} {\bfseries 90} (2003) 131802}, \href{http://arxiv.org/abs/hep-ex/0212003}{{\ttfamily arXiv:hep-ex/0212003}}.

\bibitem{Colaresi:2022obx}
J.~Colaresi, J.~I. Collar, T.~W. Hossbach, C.~M. Lewis, and K.~M. Yocum, ``{Measurement of Coherent Elastic Neutrino-Nucleus Scattering from Reactor Antineutrinos},'' \href{http://dx.doi.org/10.1103/PhysRevLett.129.211802}{{\em Phys. Rev. Lett.} {\bfseries 129} no.~21, (2022) 211802}, \href{http://arxiv.org/abs/2202.09672}{{\ttfamily arXiv:2202.09672 [hep-ex]}}.

\bibitem{Chen:2014dsa}
J.-W. Chen, H.-C. Chi, H.-B. Li, C.~P. Liu, L.~Singh, H.~T. Wong, C.-L. Wu, and C.-P. Wu, ``{Constraints on millicharged neutrinos via analysis of data from atomic ionizations with germanium detectors at sub-keV sensitivities},'' \href{http://dx.doi.org/10.1103/PhysRevD.90.011301}{{\em Phys. Rev. D} {\bfseries 90} no.~1, (2014) 011301}, \href{http://arxiv.org/abs/1405.7168}{{\ttfamily arXiv:1405.7168 [hep-ph]}}.

\bibitem{AtzoriCorona:2022qrf}
M.~Atzori~Corona, M.~Cadeddu, N.~Cargioli, F.~Dordei, C.~Giunti, Y.~F. Li, C.~A. Ternes, and Y.~Y. Zhang, ``{Impact of the Dresden-II and COHERENT neutrino scattering data on neutrino electromagnetic properties and electroweak physics},'' \href{http://dx.doi.org/10.1007/JHEP09(2022)164}{{\em JHEP} {\bfseries 09} (2022) 164}, \href{http://arxiv.org/abs/2205.09484}{{\ttfamily arXiv:2205.09484 [hep-ph]}}.

\bibitem{COHERENT:2020iec}
{\bfseries COHERENT} Collaboration, D.~Akimov {\em et~al.}, ``{First Measurement of Coherent Elastic Neutrino-Nucleus Scattering on Argon},'' \href{http://dx.doi.org/10.1103/PhysRevLett.126.012002}{{\em Phys. Rev. Lett.} {\bfseries 126} no.~1, (2021) 012002}, \href{http://arxiv.org/abs/2003.10630}{{\ttfamily arXiv:2003.10630 [nucl-ex]}}.

\bibitem{COHERENT:2021xmm}
{\bfseries COHERENT} Collaboration, D.~Akimov {\em et~al.}, ``{Measurement of the Coherent Elastic Neutrino-Nucleus Scattering Cross Section on CsI by COHERENT},'' \href{http://dx.doi.org/10.1103/PhysRevLett.129.081801}{{\em Phys. Rev. Lett.} {\bfseries 129} no.~8, (2022) 081801}, \href{http://arxiv.org/abs/2110.07730}{{\ttfamily arXiv:2110.07730 [hep-ex]}}.

\bibitem{XMASS:2020zke}
{\bfseries XMASS} Collaboration, K.~Abe {\em et~al.}, ``{Search for exotic neutrino-electron interactions using solar neutrinos in XMASS-I},'' \href{http://dx.doi.org/10.1016/j.physletb.2020.135741}{{\em Phys. Lett. B} {\bfseries 809} (2020) 135741}, \href{http://arxiv.org/abs/2005.11891}{{\ttfamily arXiv:2005.11891 [hep-ex]}}.

\bibitem{LZ:2022lsv}
{\bfseries LZ} Collaboration, J.~Aalbers {\em et~al.}, ``{First Dark Matter Search Results from the LUX-ZEPLIN (LZ) Experiment},'' \href{http://dx.doi.org/10.1103/PhysRevLett.131.041002}{{\em Phys. Rev. Lett.} {\bfseries 131} no.~4, (2023) 041002}, \href{http://arxiv.org/abs/2207.03764}{{\ttfamily arXiv:2207.03764 [hep-ex]}}.

\bibitem{AtzoriCorona:2022jeb}
M.~Atzori~Corona, W.~M. Bonivento, M.~Cadeddu, N.~Cargioli, and F.~Dordei, ``{New constraint on neutrino magnetic moment and neutrino millicharge from LUX-ZEPLIN dark matter search results},'' \href{http://dx.doi.org/10.1103/PhysRevD.107.053001}{{\em Phys. Rev. D} {\bfseries 107} no.~5, (2023) 053001}, \href{http://arxiv.org/abs/2207.05036}{{\ttfamily arXiv:2207.05036 [hep-ph]}}.

\bibitem{XENON:2022ltv}
{\bfseries XENON} Collaboration, E.~Aprile {\em et~al.}, ``{Search for New Physics in Electronic Recoil Data from XENONnT},'' \href{http://dx.doi.org/10.1103/PhysRevLett.129.161805}{{\em Phys. Rev. Lett.} {\bfseries 129} no.~16, (2022) 161805}, \href{http://arxiv.org/abs/2207.11330}{{\ttfamily arXiv:2207.11330 [hep-ex]}}.

\bibitem{A:2022acy}
S.~K. A., A.~Majumdar, D.~K. Papoulias, H.~Prajapati, and R.~Srivastava, ``{Implications of first LZ and XENONnT results: A comparative study of neutrino properties and light mediators},'' \href{http://dx.doi.org/10.1016/j.physletb.2023.137742}{{\em Phys. Lett. B} {\bfseries 839} (2023) 137742}, \href{http://arxiv.org/abs/2208.06415}{{\ttfamily arXiv:2208.06415 [hep-ph]}}.

\bibitem{PandaX:2022ood}
{\bfseries PandaX} Collaboration, D.~Zhang {\em et~al.}, ``{Search for Light Fermionic Dark Matter Absorption on Electrons in PandaX-4T},'' \href{http://dx.doi.org/10.1103/PhysRevLett.129.161804}{{\em Phys. Rev. Lett.} {\bfseries 129} no.~16, (2022) 161804}, \href{http://arxiv.org/abs/2206.02339}{{\ttfamily arXiv:2206.02339 [hep-ex]}}.

\bibitem{Giunti:2023yha}
C.~Giunti and C.~A. Ternes, ``{Testing neutrino electromagnetic properties at current and future dark matter experiments},'' \href{http://dx.doi.org/10.1103/PhysRevD.108.095044}{{\em Phys. Rev. D} {\bfseries 108} no.~9, (2023) 095044}, \href{http://arxiv.org/abs/2309.17380}{{\ttfamily arXiv:2309.17380 [hep-ph]}}.

\bibitem{Babu:1993yh}
K.~S. Babu, T.~M. Gould, and I.~Z. Rothstein, ``{Closing the windows on MeV Tau neutrinos},'' \href{http://dx.doi.org/10.1016/0370-2693(94)90340-9}{{\em Phys. Lett. B} {\bfseries 321} (1994) 140--144}, \href{http://arxiv.org/abs/hep-ph/9310349}{{\ttfamily arXiv:hep-ph/9310349}}.

\bibitem{Gninenko:2006fi}
S.~N. Gninenko, N.~V. Krasnikov, and A.~Rubbia, ``{Search for millicharged particles in reactor neutrino experiments: A Probe of the PVLAS anomaly},'' \href{http://dx.doi.org/10.1103/PhysRevD.75.075014}{{\em Phys. Rev. D} {\bfseries 75} (2007) 075014}, \href{http://arxiv.org/abs/hep-ph/0612203}{{\ttfamily arXiv:hep-ph/0612203}}.

\bibitem{Studenikin:2013my}
A.~Studenikin, ``{New bounds on neutrino electric millicharge from limits on neutrino magnetic moment},'' \href{http://dx.doi.org/10.1209/0295-5075/107/21001}{{\em EPL} {\bfseries 107} no.~2, (2014) 21001}, \href{http://arxiv.org/abs/1302.1168}{{\ttfamily arXiv:1302.1168 [hep-ph]}}. [Erratum: EPL 107, 39901 (2014), Erratum: Europhys.Lett. 107, 39901 (2014)].

\bibitem{BEBCWA66:1986err}
{\bfseries BEBC WA66} Collaboration, H.~Grassler {\em et~al.}, ``{Prompt Neutrino Production in 400-{GeV} Proton Copper Interactions},'' \href{http://dx.doi.org/10.1016/0550-3213(86)90246-4}{{\em Nucl. Phys. B} {\bfseries 273} (1986) 253--274}.

\bibitem{Cooper-Sarkar:1991vsl}
A.~M. Cooper-Sarkar, S.~Sarkar, J.~Guy, W.~Venus, P.~O. Hulth, and K.~Hultqvist, ``{Bound on the tau-neutrino magnetic moment from the BEBC beam dump experiment},'' \href{http://dx.doi.org/10.1016/0370-2693(92)90789-7}{{\em Phys. Lett. B} {\bfseries 280} (1992) 153--158}.

\bibitem{Beda:2012zz}
A.~G. Beda, V.~B. Brudanin, V.~G. Egorov, D.~V. Medvedev, V.~S. Pogosov, M.~V. Shirchenko, and A.~S. Starostin, ``{The results of search for the neutrino magnetic moment in GEMMA experiment},'' \href{http://dx.doi.org/10.1155/2012/350150}{{\em Adv. High Energy Phys.} {\bfseries 2012} (2012) 350150}.

\bibitem{CONUS:2022qbb}
{\bfseries CONUS} Collaboration, H.~Bonet {\em et~al.}, ``{First upper limits on neutrino electromagnetic properties from the CONUS experiment},'' \href{http://dx.doi.org/10.1140/epjc/s10052-022-10722-1}{{\em Eur. Phys. J. C} {\bfseries 82} no.~9, (2022) 813}, \href{http://arxiv.org/abs/2201.12257}{{\ttfamily arXiv:2201.12257 [hep-ex]}}.

\bibitem{LSND:2001akn}
{\bfseries LSND} Collaboration, L.~B. Auerbach {\em et~al.}, ``{Measurement of electron - neutrino - electron elastic scattering},'' \href{http://dx.doi.org/10.1103/PhysRevD.63.112001}{{\em Phys. Rev. D} {\bfseries 63} (2001) 112001}, \href{http://arxiv.org/abs/hep-ex/0101039}{{\ttfamily arXiv:hep-ex/0101039}}.

\bibitem{Das:2020egb}
A.~Das, D.~Ghosh, C.~Giunti, and A.~Thalapillil, ``{Neutrino charge constraints from scattering to the weak gravity conjecture to neutron stars},'' \href{http://dx.doi.org/10.1103/PhysRevD.102.115009}{{\em Phys. Rev. D} {\bfseries 102} no.~11, (2020) 115009}, \href{http://arxiv.org/abs/2005.12304}{{\ttfamily arXiv:2005.12304 [hep-ph]}}.

\bibitem{DONUT:2001zvi}
{\bfseries DONUT} Collaboration, R.~Schwienhorst {\em et~al.}, ``{A New upper limit for the tau - neutrino magnetic moment},'' \href{http://dx.doi.org/10.1016/S0370-2693(01)00746-8}{{\em Phys. Lett. B} {\bfseries 513} (2001) 23--29}, \href{http://arxiv.org/abs/hep-ex/0102026}{{\ttfamily arXiv:hep-ex/0102026}}.

\bibitem{PhysRevD.37.3107}
J.~Baumann, R.~G\"ahler, J.~Kalus, and W.~Mampe, ``Experimental limit for the charge of the free neutron,'' \href{http://dx.doi.org/10.1103/PhysRevD.37.3107}{{\em Phys. Rev. D} {\bfseries 37} (Jun, 1988) 3107--3112}. \url{https://link.aps.org/doi/10.1103/PhysRevD.37.3107}.

\bibitem{PhysRevA.83.052101}
G.~Bressi, G.~Carugno, F.~Della~Valle, G.~Galeazzi, G.~Ruoso, and G.~Sartori, ``Testing the neutrality of matter by acoustic means in a spherical resonator,'' \href{http://dx.doi.org/10.1103/PhysRevA.83.052101}{{\em Phys. Rev. A} {\bfseries 83} (May, 2011) 052101}. \url{https://link.aps.org/doi/10.1103/PhysRevA.83.052101}.

\bibitem{Muong-2:2023cdq}
{\bfseries Muon g-2} Collaboration, D.~P. Aguillard {\em et~al.}, ``{Measurement of the Positive Muon Anomalous Magnetic Moment to 0.20~ppm},'' \href{http://dx.doi.org/10.1103/PhysRevLett.131.161802}{{\em Phys. Rev. Lett.} {\bfseries 131} no.~16, (2023) 161802}, \href{http://arxiv.org/abs/2308.06230}{{\ttfamily arXiv:2308.06230 [hep-ex]}}.

\bibitem{ParticleDataGroup:2024cfk}
{\bfseries Particle Data Group} Collaboration, S.~Navas {\em et~al.}, ``{Review of particle physics},'' \href{http://dx.doi.org/10.1103/PhysRevD.110.030001}{{\em Phys. Rev. D} {\bfseries 110} no.~3, (2024) 030001}.

\bibitem{Fung:2023euv}
A.~Fung, S.~Heeba, Q.~Liu, V.~Muralidharan, K.~Schutz, and A.~C. Vincent, ``{New bounds on light millicharged particles from the tip of the red-giant branch},'' \href{http://dx.doi.org/10.1103/PhysRevD.109.083011}{{\em Phys. Rev. D} {\bfseries 109} no.~8, (2024) 083011}, \href{http://arxiv.org/abs/2309.06465}{{\ttfamily arXiv:2309.06465 [hep-ph]}}.

\bibitem{Raffelt:1999gv}
G.~G. Raffelt, ``{Limits on neutrino electromagnetic properties: An update},'' \href{http://dx.doi.org/10.1016/S0370-1573(99)00074-5}{{\em Phys. Rept.} {\bfseries 320} (1999) 319--327}.

\bibitem{Barbiellini:1987zz}
G.~Barbiellini and G.~Cocconi, ``{Electric Charge of the Neutrinos from SN1987A},'' \href{http://dx.doi.org/10.1038/329021b0}{{\em Nature} {\bfseries 329} (1987) 21--22}.

\bibitem{Kamiokande-II:1987idp}
{\bfseries Kamiokande-II} Collaboration, K.~Hirata {\em et~al.}, ``{Observation of a Neutrino Burst from the Supernova SN 1987a},'' \href{http://dx.doi.org/10.1103/PhysRevLett.58.1490}{{\em Phys. Rev. Lett.} {\bfseries 58} (1987) 1490--1493}.

\bibitem{Studenikin:2012vi}
A.~I. Studenikin and I.~Tokarev, ``{Millicharged neutrino with anomalous magnetic moment in rotating magnetized matter},'' \href{http://dx.doi.org/10.1016/j.nuclphysb.2014.04.026}{{\em Nucl. Phys. B} {\bfseries 884} (2014) 396--407}, \href{http://arxiv.org/abs/1209.3245}{{\ttfamily arXiv:1209.3245 [hep-ph]}}.

\bibitem{Mathur:2021trm}
V.~Mathur, I.~M. Shoemaker, and Z.~Tabrizi, ``{Using DUNE to shed light on the electromagnetic properties of neutrinos},'' \href{http://dx.doi.org/10.1007/JHEP10(2022)041}{{\em JHEP} {\bfseries 10} (2022) 041}, \href{http://arxiv.org/abs/2111.14884}{{\ttfamily arXiv:2111.14884 [hep-ph]}}.

\bibitem{MammenAbraham:2023psg}
R.~Mammen~Abraham, S.~Foroughi-Abari, F.~Kling, and Y.-D. Tsai, ``{Neutrino electromagnetic properties and the weak mixing angle at the LHC Forward Physics Facility},'' \href{http://dx.doi.org/10.1103/PhysRevD.111.015029}{{\em Phys. Rev. D} {\bfseries 111} no.~1, (2025) 015029}, \href{http://arxiv.org/abs/2301.10254}{{\ttfamily arXiv:2301.10254 [hep-ph]}}.

\bibitem{Ismail:2021dyp}
A.~Ismail, S.~Jana, and R.~M. Abraham, ``{Neutrino up-scattering via the dipole portal at forward LHC detectors},'' \href{http://dx.doi.org/10.1103/PhysRevD.105.055008}{{\em Phys. Rev. D} {\bfseries 105} no.~5, (2022) 055008}, \href{http://arxiv.org/abs/2109.05032}{{\ttfamily arXiv:2109.05032 [hep-ph]}}.

\bibitem{Huang:2021mki}
G.-y. Huang, S.~Jana, M.~Lindner, and W.~Rodejohann, ``{Probing new physics at future tau neutrino telescopes},'' \href{http://dx.doi.org/10.1088/1475-7516/2022/02/038}{{\em JCAP} {\bfseries 02} no.~02, (2022) 038}, \href{http://arxiv.org/abs/2112.09476}{{\ttfamily arXiv:2112.09476 [hep-ph]}}.

\bibitem{Huang:2022pce}
G.-y. Huang, S.~Jana, M.~Lindner, and W.~Rodejohann, ``{Probing heavy sterile neutrinos at neutrino telescopes via the dipole portal},'' \href{http://dx.doi.org/10.1016/j.physletb.2023.137842}{{\em Phys. Lett. B} {\bfseries 840} (2023) 137842}, \href{http://arxiv.org/abs/2204.10347}{{\ttfamily arXiv:2204.10347 [hep-ph]}}.

\bibitem{Jana:2022tsa}
S.~Jana, Y.~P. Porto-Silva, and M.~Sen, ``{Exploiting a future galactic supernova to probe neutrino magnetic moments},'' \href{http://dx.doi.org/10.1088/1475-7516/2022/09/079}{{\em JCAP} {\bfseries 09} (2022) 079}, \href{http://arxiv.org/abs/2203.01950}{{\ttfamily arXiv:2203.01950 [hep-ph]}}.

\bibitem{Jana:2023ufy}
S.~Jana and Y.~Porto, ``{Resonances of Supernova Neutrinos in Twisting Magnetic Fields},'' \href{http://dx.doi.org/10.1103/PhysRevLett.132.101005}{{\em Phys. Rev. Lett.} {\bfseries 132} no.~10, (2024) 101005}, \href{http://arxiv.org/abs/2303.13572}{{\ttfamily arXiv:2303.13572 [hep-ph]}}.

\bibitem{Maruno:1991vr}
M.~Maruno, E.~Takasugi, and M.~Tanaka, ``{Minicharge nonconservation in SU(3)(C) x SU(2)(L) x U(1)(Y) models},'' \href{http://dx.doi.org/10.1143/PTP.86.907}{{\em Prog. Theor. Phys.} {\bfseries 86} (1991) 907--916}.

\bibitem{Takasugi:1991ai}
E.~Takasugi and M.~Tanaka, ``{Charge nonconservation and charges of neutrinos, neutron and atoms},'' \href{http://dx.doi.org/10.1103/PhysRevD.44.3706}{{\em Phys. Rev. D} {\bfseries 44} (1991) 3706--3708}.

\bibitem{Takasugi:1991wa}
E.~Takasugi and M.~Tanaka, ``{Charged neutrinos and atoms in the standard model},'' \href{http://dx.doi.org/10.1143/PTP.87.679}{{\em Prog. Theor. Phys.} {\bfseries 87} (1992) 679--684}.

\bibitem{Ignatiev:1996np}
A.~Y. Ignatiev and G.~C. Joshi, ``{Possible electric charge nonconservation and dequantization in SU(2) x U(1) models with hard symmetry breaking},'' \href{http://dx.doi.org/10.1016/0370-2693(96)00569-2}{{\em Phys. Lett. B} {\bfseries 381} (1996) 216--220}, \href{http://arxiv.org/abs/hep-ph/9604238}{{\ttfamily arXiv:hep-ph/9604238}}.

\bibitem{Ignatiev:1997pk}
A.~Y. Ignatiev and G.~C. Joshi, ``{Hybrid SU(2) x U(1) models, electric charge nonconservation and the photon mass},'' \href{http://dx.doi.org/10.1016/S0960-0779(98)00232-X}{{\em Chaos Solitons Fractals} {\bfseries 10} (1999) 1}, \href{http://arxiv.org/abs/hep-ph/9710552}{{\ttfamily arXiv:hep-ph/9710552}}.

\bibitem{Babu:1990vw}
K.~S. Babu and R.~N. Mohapatra, ``{WHY DOES ELECTROMAGNETISM CONSERVE PARITY?},'' \href{http://dx.doi.org/10.1103/PhysRevD.42.3866}{{\em Phys. Rev. D} {\bfseries 42} (1990) 3866--3869}.

\end{thebibliography}\endgroup
\end{document}